\documentclass[aip, apl, preprint, superscriptaddress]{revtex4-2}

\usepackage{graphicx}
\usepackage{dcolumn}
\usepackage{bm}
\usepackage{upgreek}
\usepackage{color,soul}
\usepackage{amsmath}

\begin{document}

\title{Spin-wave propagation at low temperatures in YIG thin films on YSGG substrates}

\author{José Elias Abrão}
\affiliation{NanoSpin, Department of Applied Physics, Aalto University School of Science, P.O. Box 15100, FI-00076 Aalto, Finland}
\author{Daan Weltens}
\affiliation{NanoSpin, Department of Applied Physics, Aalto University School of Science, P.O. Box 15100, FI-00076 Aalto, Finland}
\author{Rhodri Mansell}
\email[Corresponding author: ] {rhodri.mansell@aalto.fi}
\affiliation{NanoSpin, Department of Applied Physics, Aalto University School of Science, P.O. Box 15100, FI-00076 Aalto, Finland}
\author{Sebastiaan van Dijken}
\email[Corresponding author: ]{sebastiaan.van.dijken@aalto.fi}
\affiliation{NanoSpin, Department of Applied Physics, Aalto University School of Science, P.O. Box 15100, FI-00076 Aalto, Finland}
\author{Luk{\'{a}}{\v{s}} Flaj{\v{s}}man}
\email[Corresponding author: ] {lukas.flajsman@aalto.fi}
\affiliation{NanoSpin, Department of Applied Physics, Aalto University School of Science, P.O. Box 15100, FI-00076 Aalto, Finland}

\begin{abstract}
The use of spin waves in magnetic thin films at cryogenic temperatures has long been hindered by the lack of a suitable material platform. Yttrium iron garnet (YIG) is the leading candidate, yet it is typically grown on gadolinium gallium garnet (GGG) substrates, which develop a large paramagnetic moment at low temperatures. This substrate effect limits spin-wave propagation. In this work, we demonstrate that thin YIG films grown on yttrium scandium gallium garnet (YSGG) substrates support robust spin-wave propagation in the Damon-Eshbach geometry, measurable down to 2 K under applied magnetic fields up to 150 mT. Compared with YIG/GGG, YIG/YSGG films exhibit narrower ferromagnetic resonance (FMR) linewidths at low temperatures and are free from the atomic interdiffusion effects that degrade the performance of YIG/GGG systems. These results establish YIG/YSGG thin films as a promising low-temperature platform, overcoming the intrinsic limitations of YIG/GGG and opening new opportunities for scalable magnonic and hybrid quantum devices operating under cryogenic conditions.
\end{abstract}

\maketitle



\begin{figure*}[htbp]
    \centering
    \includegraphics[width=1.0\linewidth]{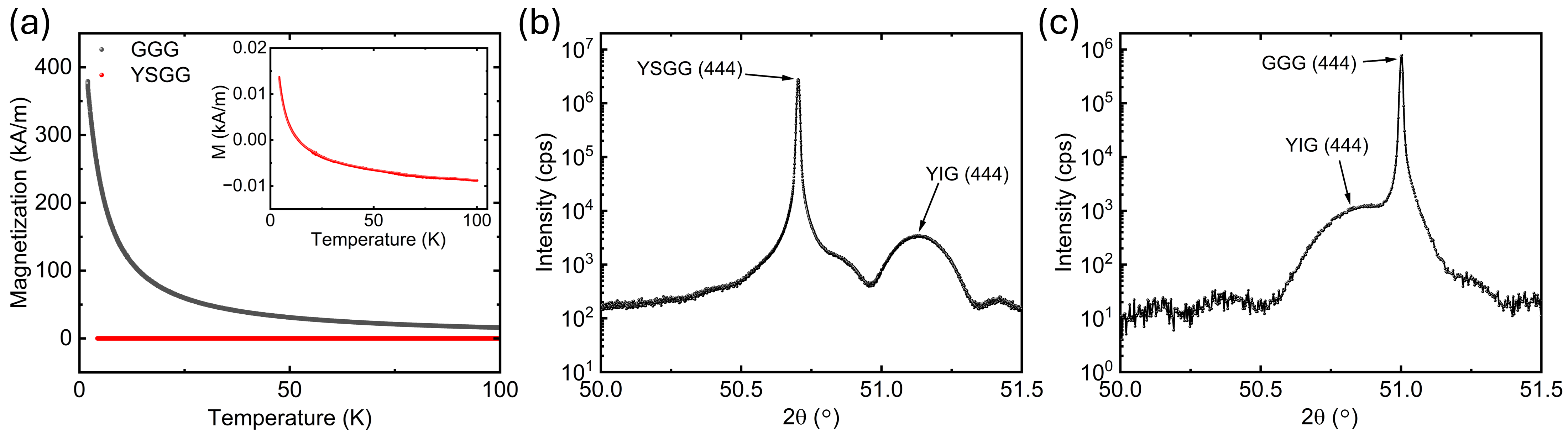}
    \caption{(a) Temperature dependence of the magnetization of GGG (black) and YSGG (red) substrates measured under a 1 T applied field in the range $4-100$ K. The inset shows the YSGG data with an expanded y-axis scale. (b) XRD $\theta$-$2\theta$ scan of the YIG/YSGG film. (c) XRD $\theta$-$2\theta$ scan of the YIG/GGG film. The (444) reflections from the film and substrate are labeled.}
    \label{fig1}
\end{figure*}

Propagating spin waves in yttrium iron garnet (YIG) enable long-range, low-loss information transport at room temperature \cite{Chumak2014,Kuznetsov2025,Flebus2024}. Extending this functionality to cryogenic temperatures could open new avenues for hybrid quantum systems \cite{YUAN2022,Lachance-Quirion2019,Pal2024}. While quantum magnonics has advanced rapidly in recent years, most demonstrations rely on YIG spheres \cite{Lachance-Quirion2020,Xu2023} or micron-thick films \cite{Baity2021}, which limit scalability and on-chip integration. Single-crystal YIG films grown on gadolinium gallium garnet (GGG) substrates show low FMR linewidths \cite{Qin2018,Youssef2025,WillCole2023} and support efficient spin-wave propagation at room temperature \cite{Kuznetsov2025,Qin2018}. However, the temperature dependence of the FMR linewidth varies markedly between films \cite{Haidar2015, Jermain2017,WillCole2023,Guo2022,Youssef2025,Knauer2023} with many showing strong increases in linewidth at lower temperatures \cite{Jermain2017,Guo2022, Youssef2025,Knauer2023}. In addition, YIG/GGG films at low temperatures show a strong suppression of spin-wave transport \cite{Schmoll2025,Knauer2023}. These effects are commonly attributed to the large paramagnetic moment of GGG at low temperatures, which generates inhomogeneous stray magnetic fields \cite{Serha2024,SERHA2025,Schmoll2025}, as well as to Gd diffusion from the substrate that enhances impurity-related relaxation mechanisms \cite{Mitra2017,Jermain2017,WillCole2023}. Two main strategies have been pursued to mitigate these issues. The first involves mechanically thinning or completely removing the GGG substrate \cite{Kosen2019,Xu2025}. The second focuses on employing alternative garnet substrates or buffer layers, such as Y$_3$Sc$_2$Ga$_3$O$_{12}$ (YSGG) \cite{Legrand2025,Guo2023,Youssef2025}, solid-solution garnets based on Y$_3$Sc$_2$Ga$_3$O$_{12}$-Y$_3$Sc$_2$Al$_3$O$_{12}$ and Y$_3$Sc$_2$Ga$_3$O$_{12}$-Y$_3$Al$_5$O$_{12}$ (YSGAG) \cite{Guguschev2025}, and Y$_3$Sc$_{2.5}$Al$_{2.5}$O$_{12}$ (YSAG) buffer layers on GGG \cite{Guo2022}.

\begin{figure}[htbp]
    \centering
    \includegraphics[width=0.9\linewidth]{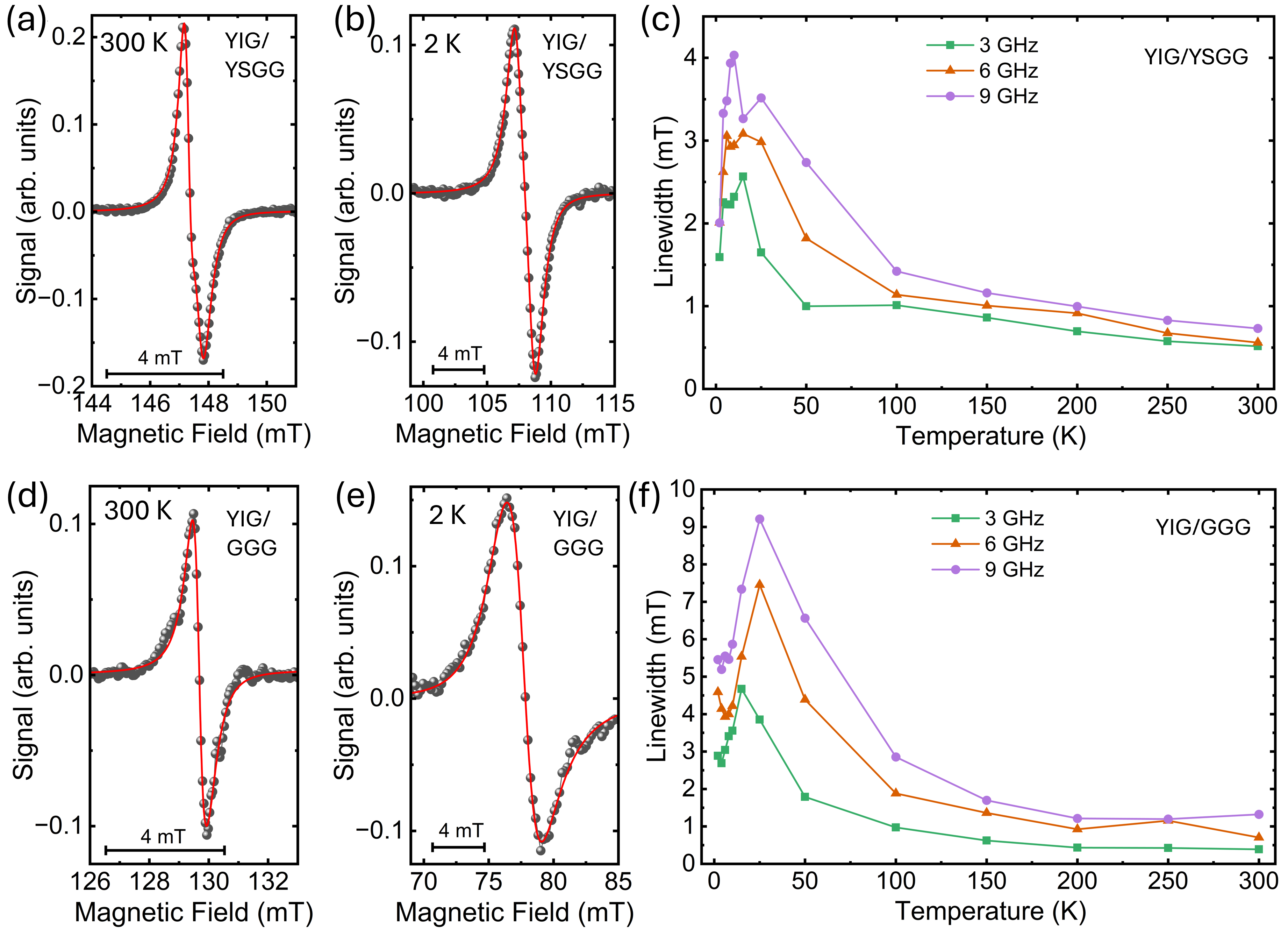}
    \caption{(a) FMR signal of the YIG/YSGG film measured at 6 GHz and 300 K. The black dots represent the experimental data, and the red line shows the fitted curve. (b)  FMR signal of the YIG/YSGG film measured at 6 GHz and 2 K. (c) Temperature dependence of the extracted FMR linewidths for the YIG/YSGG film at different frequencies. (d) FMR signal of the YIG/GGG film measured at 6 GHz and 300 K. (e) FMR signal of the YIG/GGG film measured at 6 GHz and 2 K. (f) Temperature dependence of the extracted FMR linewidths for the YIG/GGG film. Note the different y-axis scales in (c) and (f). All FMR spectra were fitted using two differentiated Lorentzian lineshapes, and the linewidths in (c) and (f) correspond to the higher-amplitude Lorentzian component.}
    \label{fig2}
\end{figure}

YSGG provides a good lattice match to YIG without exhibiting strong paramagnetism. Previous studies have shown that YIG/YSGG films exhibit low damping in low-temperature FMR measurements \cite{Legrand2025,Guo2023,Youssef2025}, but spin-wave propagation in this system has not yet been reported. Here, we grow high-quality YIG thin films on YSGG substrates by pulsed laser deposition (PLD) and demonstrate robust spin-wave propagation down to 2 K under applied magnetic fields up to 150 mT. These results establish YIG/YSGG as a suitable materials platform for low-temperature magnonic devices.

Figure 1(a) compares the magnetic response of YSGG and GGG substrates measured by vibrating sample magnetometry under a 1 T field between 4 K and 100 K. The GGG substrate exhibits a large paramagnetic moment that increases with decreasing temperature, following the Curie-Weiss law. In contrast, YSGG shows no detectable signal on this scale. The inset reveals that YSGG undergoes a transition from diamagnetic to weakly paramagnetic behavior at low temperatures, likely due to residual impurities.

The YIG/YSGG film was grown on a commercially supplied (111)-oriented YSGG single-crystal substrate using a TSST PLD system. Prior to growth, the substrate was annealed ex-situ at 1000$^\circ$C in oxygen at atmospheric pressure for four hours to ensure a clean, well-ordered surface. Growth was carried out at 875 $^\circ$C in 2.5$\times10^{-3}$ mbar O$_2$ using a KrF laser (248 nm) with a fluence of 2.5 mJ/cm$^2$ and a 7 Hz repetition rate. After deposition, the O$_2$ pressure was increased to $5\times10^{-3}$ mbar, and the sample was cooled to room temperature at 1 $^\circ$C/min. X-ray reflectivity (XRR) measurements determined the film thickness to be 52 nm. The X-ray diffraction (XRD) $\theta$-$2\theta$ scan in Fig.\ 1(b) shows that the YIG film on YSGG is under tensile strain. For comparison, a YIG film was grown on a GGG substrate under identical conditions, with the corresponding XRD spectrum shown in Fig.\ 1(c). In this case, the YIG peak appears as a shoulder on the low-angle side of the substrate peak, indicating a slight compressive strain from the GGG substrate. 

Figure 2(a) presents the room-temperature FMR spectrum of the YIG/YSGG film at 6 GHz, measured using a CryoFMR setup (NanOsc) installed into a Quantum Design DynaCool PPMS system. This technique applies a small oscillating magnetic field and detects the derivative of the Lorentzian FMR signal via lock-in detection (see the supplementary material for details on the measurement and fitting procedure). Figure 2(b) shows the corresponding signal at 2 K, where the resonance exhibits a broader linewidth. Figure 2(c) summarizes the temperature dependence from 2 K to 300 K of the extracted FMR linewidths for the YIG/YSGG film measured at 3 GHz, 6 GHz, and 9 GHz. Figures 2(d) and 2(e) display the 6 GHz FMR spectra for the YIG/GGG film at 300 K and 2 K, respectively. At 300 K, the linewidth is comparable to that of the YIG/YSGG film, but at 2 K the linewidth is considerably broader than both the room-temperature result and the YIG/YSGG film at 2 K. Figure 2(f) shows the temperature dependence of the linewidth for the YIG/GGG film. Comparing Figs.\ 2(c) and 2(f) reveals that both films exhibit a pronounced linewidth peak near 25 K, followed by a reduction in linewidth as the temperature decreases to 2 K. The peak linewidth in YIG/GGG is approximately twice that of YIG/YSGG, and this twofold difference persists down to the lowest temperatures. A similar linewidth peak has been previously reported for YIG/GGG films \cite{Jermain2017,WillCole2023,Legrand2025} and attributed to impurity-related relaxation mechanisms \cite{Jermain2017,WillCole2023}, caused by rare-earth impurities \cite{Seiden1964}. Since both films were grown under identical conditions, impurities originating from the PLD target likely contribute to this feature in both cases. However, the consistently broader linewidth in YIG/GGG suggests an additional contribution from Gd diffusion out of the GGG substrate \cite{WillCole2023,Mitra2017}. The temperature dependence of the saturation magnetization, Gilbert damping parameter, inhomogeneous linewidth broadening, and out-of-plane anisotropy extracted from the FMR fits for both films are provided in the supplementary material.

\begin{figure*}[htbp]
    \centering
    \includegraphics[width=1.0\linewidth]{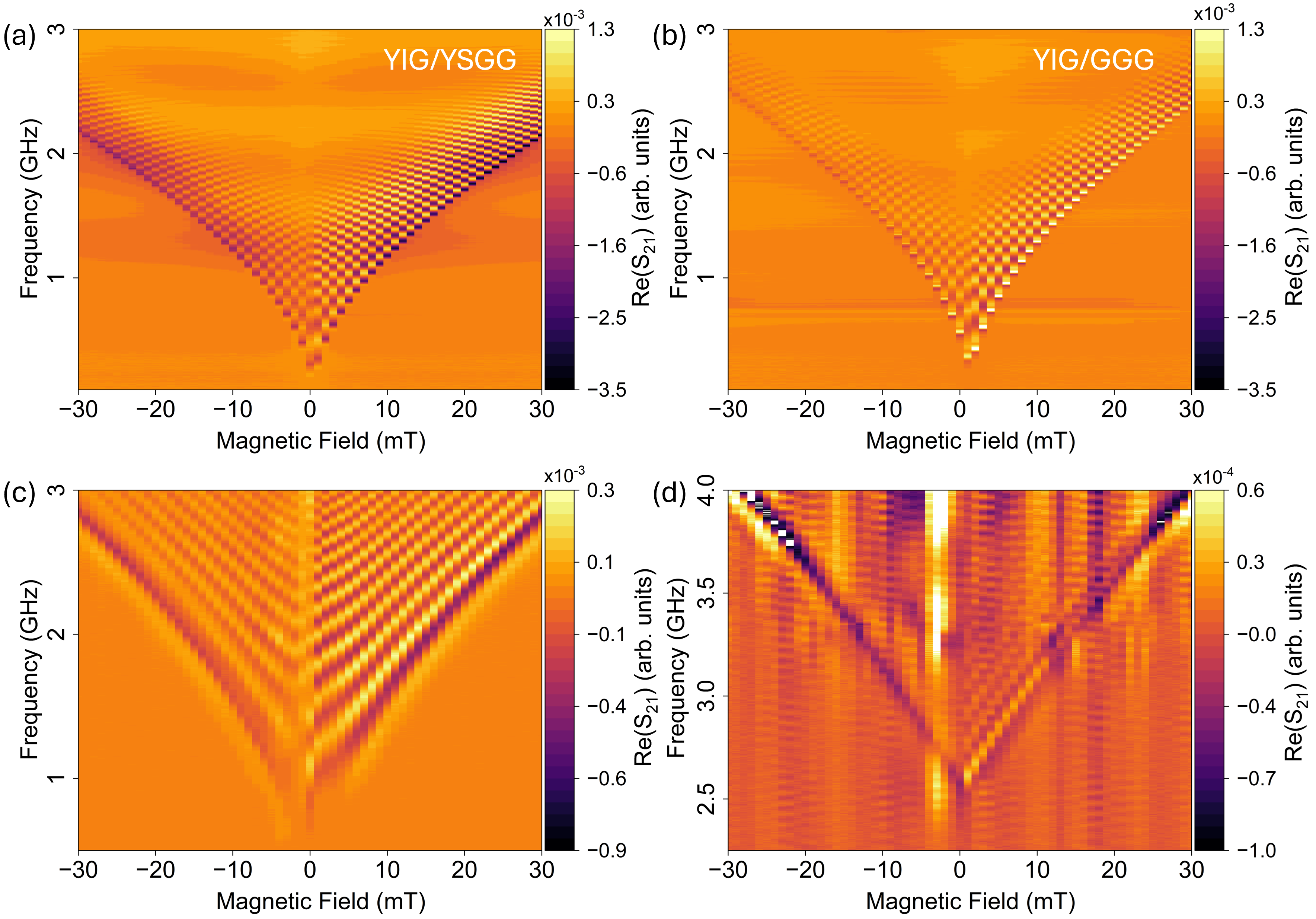}
    \caption{(a) Heat map of the real part of S$_{21}$ as a function of frequency and applied magnetic field for the YIG/YSGG film at 300 K. (b) Same as (a) for the YIG/GGG sample at 300 K. (c) Same as (a) for the YIG/YSGG sample at 2 K. (d) Same as (a) for the YIG/GGG sample at 2 K. In all measurements, the excitation power was set to $-15$ dBm, using parallel antennas separated by 10 $\upmu$m. A reference spectrum acquired at higher magnetic fields was subtracted from the raw data to obtain the S$_{21}$ signal. Additional background subtraction was applied in (c) and (d) to remove non-magnetic artifacts.}
    \label{fig3}
\end{figure*}

Previous studies have shown that spin-wave propagation is possible in thin-film YIG/GGG samples at low temperatures, but that propagation is strongly affected at high fields by inhomogeneous stray fields arising from the paramagnetic GGG substrate \cite{Serha2024, Schmoll2025, Knauer2023}. Spatial variations in the local effective field can suppress spin-wave transport even in samples with relatively low damping \cite{Das2023}. To characterize spin-wave propagation in both YIG/YSGG and YIG/GGG films, we patterned microwave antennas onto the samples using photolithography, followed by magnetron sputtering of Ta (5 nm) / Au (120 nm) and lift-off. Antennas with spacings of 10 $\upmu$m, 20 $\upmu$m, and 50 $\upmu$m were fabricated and mounted on a custom holder integrated with the CryoFMR sample rod inside the PPMS. The antennas were connected to a Copper Mountain Technologies C1209 2-port VNA, operated with an input power of $-15$ dBm, to extract the S$_{21}$ scattering parameters. Figure\ 3(a) shows the real part of the S$_{21}$ spectra measured on YIG/YSGG at 300 K with the magnetic field swept from $-30$ mT to 30 mT in the Damon-Eshbach configuration using 10 $\upmu$m antenna spacing. Clear signatures of propagating spin waves are observed \cite{Yu2014,Qin2018}, with the signal shifting to higher frequencies as the applied field increases, consistent with theory. The asymmetry in spin-wave amplitude between positive and negative fields arises from the intrinsic asymmetry in antenna excitation efficiency \cite{Schneider2008}. Equivalent measurements on YIG/GGG at 300 K, shown in Fig.\ 3(b), also show clear spin-wave propagation under identical conditions. At cryogenic temperatures, however, a stark contrast emerges between the two systems. Figure\ 3(c) displays the data for YIG/YSGG at 2 K, where pronounced oscillations in the S$_{21}$ signal indicate robust spin-wave transport between the antennas. In contrast, the YIG/GGG film at 2 K exhibits a strong suppression of spin-wave propagation. Only weak signals --- about an order of magnitude smaller than in YIG/YSGG --- are detectable, and these are confined to the positive-field branch up to approximately 20 mT. The group velocities of the propagating spin waves, extracted from the oscillations in Re(S$_{21}$) for the data shown in Fig.\ 3, are presented in the supplementary material.

Figure\ 4 presents line scans of the real part of the S$_{21}$ signal as a function of frequency for the YIG/YSGG sample at 2 K and various antenna spacings. At 5 mT, clear oscillations are visible for a 10 $\upmu$m spacing (Fig.\ 4(a)), confirming robust spin-wave propagation. The corresponding spectra for 20 µm and 50 µm spacings are shown in Figs.\ 4(b) and 4(c), respectively. While the 20 $\upmu$m signal remains comparable in magnitude to that at 10 µm, the 50 $\upmu$m spacing yields a reduction of roughly an order of magnitude. This attenuation likely reflects spatial inhomogeneities in the film, possibly originating from the PLD growth process. Figures\ 4(d)-(f) show the S$_{21}$ response for the 10 $\upmu$m spacing under higher applied fields. Clear spin-wave propagation persists at 50 mT, 100 mT, and 150 mT. In each case, a distinct FMR-like feature precedes the onset of propagation, which we attribute to remote FMR excitation induced by the drive antenna beneath the detection antenna --- a known effect that becomes more pronounced at small antenna spacings \cite{Sushruth2020}. Importantly, the spin-wave spectroscopy data in Figs.\ 3 and 4 demonstrate robust, field-tunable spin-wave transport at temperatures and frequencies relevant to quantum magnonics \cite{Lachance-Quirion2020}.

\begin{figure*}[htbp]
    \centering
    \includegraphics[width=1.0\linewidth]{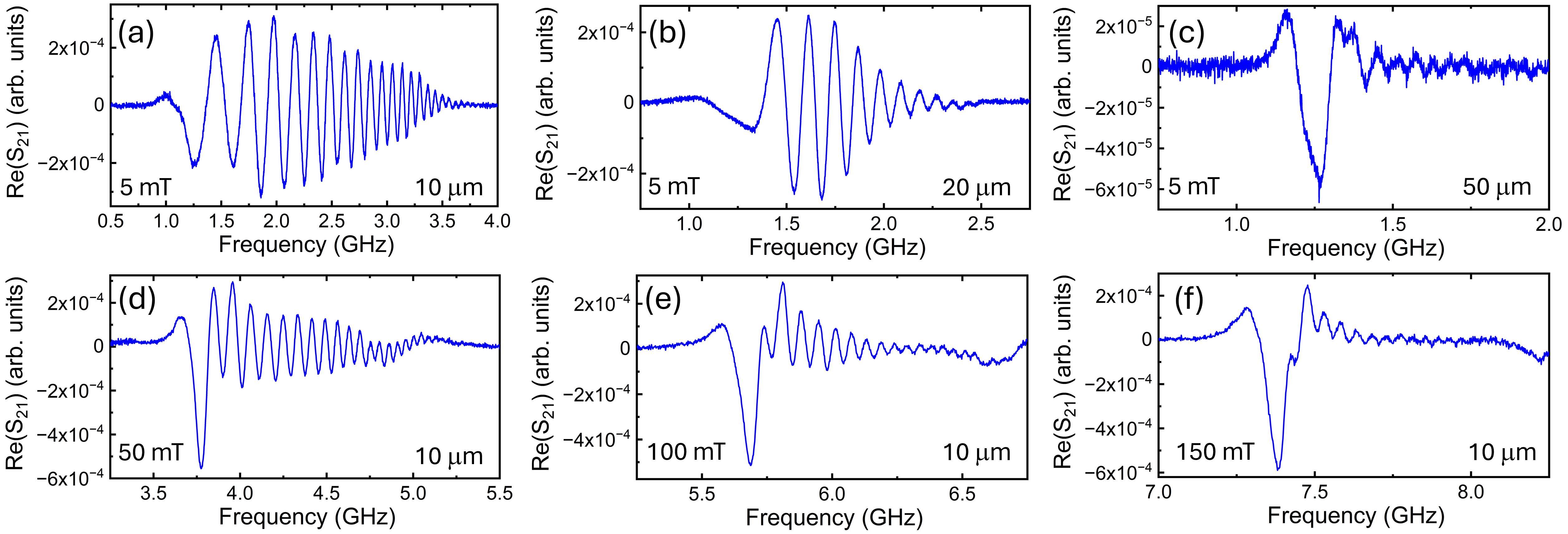}
    \caption{(a) Real part of S$_{21}$ as a function of frequency at 2 K and 5 mT, measured with parallel antennas separated by 10 $\upmu$m. (b) Same as (a) with 20 $\upmu$m antenna spacing. (c) Same as (a) with 50 $\upmu$m spacing. (d-f) Same as (a) with 10 $\upmu$m spacing at applied fields of 50 mT, 100 mT, and 150 mT, respectively. In all cases, a reference measurement at higher applied magnetic fields was subtracted from the raw data to obtain the S$_{21}$ signal, and additional background subtraction was applied to remove non-magnetic artifacts.}
    \label{fig4}
\end{figure*}

In summary, high-quality YIG thin films have been successfully grown on YSGG substrates by PLD, exhibiting narrow FMR linewidths and robust spin-wave propagation down to 2 K. The absence of strong paramagnetism and Gd diffusion in YSGG prevents the degradation seen in YIG/GGG films at low temperatures, enabling coherent spin-wave transport in the quantum-relevant regime. These results establish YIG/YSGG thin films as a scalable materials platform for cryogenic magnonics, with promising applications in on-chip magnonic interconnects, quantum transducers, and hybrid quantum systems.

\section*{Supplementary material}
The supplementary material contains a description of the FMR data fitting procedure, along with the extraction of the saturation magnetization, Gilbert damping parameter, inhomogeneous linewidth broadening, and out-of-plane anisotropy, all as a function of temperature, from the FMR data. It further contains the spin-wave group velocities extracted from the data in Fig.\ 3 and atomic force microscopy (AFM) images characterizing the surface roughness of the samples.

\section*{Acknowledgments}
This work was supported by the Research Council of Finland (Grant No. 357211) and by the Finnish Quantum Flagship project (Grant No. 358877). Lithography was carried out in the OtaNano-Micronova cleanroom, supported by Aalto University. XRD was performed at the OtaNano-Nanomicroscopy Center (Aalto-NMC).  

\bibliography{YSGG}

\begin{thebibliography}{30}%
\makeatletter
\providecommand \@ifxundefined [1]{%
 \@ifx{#1\undefined}
}%
\providecommand \@ifnum [1]{%
 \ifnum #1\expandafter \@firstoftwo
 \else \expandafter \@secondoftwo
 \fi
}%
\providecommand \@ifx [1]{%
 \ifx #1\expandafter \@firstoftwo
 \else \expandafter \@secondoftwo
 \fi
}%
\providecommand \natexlab [1]{#1}%
\providecommand \enquote  [1]{``#1''}%
\providecommand \bibnamefont  [1]{#1}%
\providecommand \bibfnamefont [1]{#1}%
\providecommand \citenamefont [1]{#1}%
\providecommand \href@noop [0]{\@secondoftwo}%
\providecommand \href [0]{\begingroup \@sanitize@url \@href}%
\providecommand \@href[1]{\@@startlink{#1}\@@href}%
\providecommand \@@href[1]{\endgroup#1\@@endlink}%
\providecommand \@sanitize@url [0]{\catcode `\\12\catcode `\$12\catcode `\&12\catcode `\#12\catcode `\^12\catcode `\_12\catcode `\%12\relax}%
\providecommand \@@startlink[1]{}%
\providecommand \@@endlink[0]{}%
\providecommand \url  [0]{\begingroup\@sanitize@url \@url }%
\providecommand \@url [1]{\endgroup\@href {#1}{\urlprefix }}%
\providecommand \urlprefix  [0]{URL }%
\providecommand \Eprint [0]{\href }%
\providecommand \doibase [0]{https://doi.org/}%
\providecommand \selectlanguage [0]{\@gobble}%
\providecommand \bibinfo  [0]{\@secondoftwo}%
\providecommand \bibfield  [0]{\@secondoftwo}%
\providecommand \translation [1]{[#1]}%
\providecommand \BibitemOpen [0]{}%
\providecommand \bibitemStop [0]{}%
\providecommand \bibitemNoStop [0]{.\EOS\space}%
\providecommand \EOS [0]{\spacefactor3000\relax}%
\providecommand \BibitemShut  [1]{\csname bibitem#1\endcsname}%
\let\auto@bib@innerbib\@empty
\bibitem [{\citenamefont {Chumak}, \citenamefont {Serga},\ and\ \citenamefont {Hillebrands}(2014)}]{Chumak2014}%
  \BibitemOpen
  \bibfield  {author} {\bibinfo {author} {\bibfnamefont {A.}~\bibnamefont {Chumak}}, \bibinfo {author} {\bibfnamefont {A.}~\bibnamefont {Serga}},\ and\ \bibinfo {author} {\bibfnamefont {B.}~\bibnamefont {Hillebrands}},\ }\bibfield  {title} {\enquote {\bibinfo {title} {Magnon transistor for all-magnon data processing.}}\ }\href {https://doi.org/https://doi.org/10.1038/ncomms5700} {\bibfield  {journal} {\bibinfo  {journal} {Nat. Commun.}\ }\textbf {\bibinfo {volume} {5}},\ \bibinfo {pages} {4700} (\bibinfo {year} {2014})}\BibitemShut {NoStop}%
\bibitem [{\citenamefont {Kuznetsov}\ \emph {et~al.}(2025)\citenamefont {Kuznetsov}, \citenamefont {Qin}, \citenamefont {Flajšman},\ and\ \citenamefont {van Dijken}}]{Kuznetsov2025}%
  \BibitemOpen
  \bibfield  {author} {\bibinfo {author} {\bibfnamefont {N.}~\bibnamefont {Kuznetsov}}, \bibinfo {author} {\bibfnamefont {H.}~\bibnamefont {Qin}}, \bibinfo {author} {\bibfnamefont {L.}~\bibnamefont {Flajšman}},\ and\ \bibinfo {author} {\bibfnamefont {S.}~\bibnamefont {van Dijken}},\ }\bibfield  {title} {\enquote {\bibinfo {title} {Optical control of spin waves in hybrid magnonic-plasmonic structures},}\ }\href {https://doi.org/10.1126/sciadv.ads2420} {\bibfield  {journal} {\bibinfo  {journal} {Sci. Adv.}\ }\textbf {\bibinfo {volume} {11}},\ \bibinfo {pages} {eads2420} (\bibinfo {year} {2025})}\BibitemShut {NoStop}%
\bibitem [{\citenamefont {Flebus}\ \emph {et~al.}(2024)\citenamefont {Flebus}, \citenamefont {Grundler}, \citenamefont {Rana}, \citenamefont {Otani}, \citenamefont {Barsukov}, \citenamefont {Barman}, \citenamefont {Gubbiotti}, \citenamefont {Landeros}, \citenamefont {Akerman}, \citenamefont {Ebels}, \citenamefont {Pirro}, \citenamefont {Demidov}, \citenamefont {Schultheiss}, \citenamefont {Csaba}, \citenamefont {Wang}, \citenamefont {Ciubotaru}, \citenamefont {Nikonov}, \citenamefont {Che}, \citenamefont {Hertel}, \citenamefont {Ono}, \citenamefont {Afanasiev}, \citenamefont {Mentink}, \citenamefont {Rasing}, \citenamefont {Hillebrands}, \citenamefont {Kusminskiy}, \citenamefont {Zhang}, \citenamefont {Du}, \citenamefont {Finco}, \citenamefont {van~der Sar}, \citenamefont {Luo}, \citenamefont {Shiota}, \citenamefont {Sklenar}, \citenamefont {Yu},\ and\ \citenamefont {Rao}}]{Flebus2024}%
  \BibitemOpen
  \bibfield  {author} {\bibinfo {author} {\bibfnamefont {B.}~\bibnamefont {Flebus}}, \bibinfo {author} {\bibfnamefont {D.}~\bibnamefont {Grundler}}, \bibinfo {author} {\bibfnamefont {B.}~\bibnamefont {Rana}}, \bibinfo {author} {\bibfnamefont {Y.}~\bibnamefont {Otani}}, \bibinfo {author} {\bibfnamefont {I.}~\bibnamefont {Barsukov}}, \bibinfo {author} {\bibfnamefont {A.}~\bibnamefont {Barman}}, \bibinfo {author} {\bibfnamefont {G.}~\bibnamefont {Gubbiotti}}, \bibinfo {author} {\bibfnamefont {P.}~\bibnamefont {Landeros}}, \bibinfo {author} {\bibfnamefont {J.}~\bibnamefont {Akerman}}, \bibinfo {author} {\bibfnamefont {U.}~\bibnamefont {Ebels}}, \bibinfo {author} {\bibfnamefont {P.}~\bibnamefont {Pirro}}, \bibinfo {author} {\bibfnamefont {V.~E.}\ \bibnamefont {Demidov}}, \bibinfo {author} {\bibfnamefont {K.}~\bibnamefont {Schultheiss}}, \bibinfo {author} {\bibfnamefont {G.}~\bibnamefont {Csaba}}, \bibinfo {author} {\bibfnamefont {Q.}~\bibnamefont {Wang}}, \bibinfo {author} {\bibfnamefont {F.}~\bibnamefont
  {Ciubotaru}}, \bibinfo {author} {\bibfnamefont {D.~E.}\ \bibnamefont {Nikonov}}, \bibinfo {author} {\bibfnamefont {P.}~\bibnamefont {Che}}, \bibinfo {author} {\bibfnamefont {R.}~\bibnamefont {Hertel}}, \bibinfo {author} {\bibfnamefont {T.}~\bibnamefont {Ono}}, \bibinfo {author} {\bibfnamefont {D.}~\bibnamefont {Afanasiev}}, \bibinfo {author} {\bibfnamefont {J.}~\bibnamefont {Mentink}}, \bibinfo {author} {\bibfnamefont {T.}~\bibnamefont {Rasing}}, \bibinfo {author} {\bibfnamefont {B.}~\bibnamefont {Hillebrands}}, \bibinfo {author} {\bibfnamefont {S.~V.}\ \bibnamefont {Kusminskiy}}, \bibinfo {author} {\bibfnamefont {W.}~\bibnamefont {Zhang}}, \bibinfo {author} {\bibfnamefont {C.~R.}\ \bibnamefont {Du}}, \bibinfo {author} {\bibfnamefont {A.}~\bibnamefont {Finco}}, \bibinfo {author} {\bibfnamefont {T.}~\bibnamefont {van~der Sar}}, \bibinfo {author} {\bibfnamefont {Y.~K.}\ \bibnamefont {Luo}}, \bibinfo {author} {\bibfnamefont {Y.}~\bibnamefont {Shiota}}, \bibinfo {author} {\bibfnamefont {J.}~\bibnamefont
  {Sklenar}}, \bibinfo {author} {\bibfnamefont {T.}~\bibnamefont {Yu}},\ and\ \bibinfo {author} {\bibfnamefont {J.}~\bibnamefont {Rao}},\ }\bibfield  {title} {\enquote {\bibinfo {title} {The 2024 magnonics roadmap},}\ }\href {https://doi.org/10.1088/1361-648X/ad399c} {\bibfield  {journal} {\bibinfo  {journal} {J. Phys. Condens. Matter}\ }\textbf {\bibinfo {volume} {36}},\ \bibinfo {pages} {363501} (\bibinfo {year} {2024})}\BibitemShut {NoStop}%
\bibitem [{\citenamefont {Yuan}\ \emph {et~al.}(2022)\citenamefont {Yuan}, \citenamefont {Cao}, \citenamefont {Kamra}, \citenamefont {Duine},\ and\ \citenamefont {Yan}}]{YUAN2022}%
  \BibitemOpen
  \bibfield  {author} {\bibinfo {author} {\bibfnamefont {H.}~\bibnamefont {Yuan}}, \bibinfo {author} {\bibfnamefont {Y.}~\bibnamefont {Cao}}, \bibinfo {author} {\bibfnamefont {A.}~\bibnamefont {Kamra}}, \bibinfo {author} {\bibfnamefont {R.~A.}\ \bibnamefont {Duine}},\ and\ \bibinfo {author} {\bibfnamefont {P.}~\bibnamefont {Yan}},\ }\bibfield  {title} {\enquote {\bibinfo {title} {Quantum magnonics: When magnon spintronics meets quantum information science},}\ }\href {https://doi.org/https://doi.org/10.1016/j.physrep.2022.03.002} {\bibfield  {journal} {\bibinfo  {journal} {Phys. Rep.}\ }\textbf {\bibinfo {volume} {965}},\ \bibinfo {pages} {1--74} (\bibinfo {year} {2022})}\BibitemShut {NoStop}%
\bibitem [{\citenamefont {Lachance-Quirion}\ \emph {et~al.}(2019)\citenamefont {Lachance-Quirion}, \citenamefont {Tabuchi}, \citenamefont {Gloppe}, \citenamefont {Usami},\ and\ \citenamefont {Nakamura}}]{Lachance-Quirion2019}%
  \BibitemOpen
  \bibfield  {author} {\bibinfo {author} {\bibfnamefont {D.}~\bibnamefont {Lachance-Quirion}}, \bibinfo {author} {\bibfnamefont {Y.}~\bibnamefont {Tabuchi}}, \bibinfo {author} {\bibfnamefont {A.}~\bibnamefont {Gloppe}}, \bibinfo {author} {\bibfnamefont {K.}~\bibnamefont {Usami}},\ and\ \bibinfo {author} {\bibfnamefont {Y.}~\bibnamefont {Nakamura}},\ }\bibfield  {title} {\enquote {\bibinfo {title} {Hybrid quantum systems based on magnonics},}\ }\href {https://doi.org/10.7567/1882-0786/ab248d} {\bibfield  {journal} {\bibinfo  {journal} {Appl. Phys. Express}\ }\textbf {\bibinfo {volume} {12}},\ \bibinfo {pages} {070101} (\bibinfo {year} {2019})}\BibitemShut {NoStop}%
\bibitem [{\citenamefont {Pal}, \citenamefont {Mondal},\ and\ \citenamefont {Barman}(2024)}]{Pal2024}%
  \BibitemOpen
  \bibfield  {author} {\bibinfo {author} {\bibfnamefont {P.~K.}\ \bibnamefont {Pal}}, \bibinfo {author} {\bibfnamefont {A.~K.}\ \bibnamefont {Mondal}},\ and\ \bibinfo {author} {\bibfnamefont {A.}~\bibnamefont {Barman}},\ }\bibfield  {title} {\enquote {\bibinfo {title} {Using magnons as a quantum technology platform: a perspective},}\ }\href {https://doi.org/10.1088/1361-648X/ad6828} {\bibfield  {journal} {\bibinfo  {journal} {J. Phys. Condens. Matter}\ }\textbf {\bibinfo {volume} {36}},\ \bibinfo {pages} {441502} (\bibinfo {year} {2024})}\BibitemShut {NoStop}%
\bibitem [{\citenamefont {Lachance-Quirion}\ \emph {et~al.}(2020)\citenamefont {Lachance-Quirion}, \citenamefont {Wolski}, \citenamefont {Tabuchi}, \citenamefont {Kono}, \citenamefont {Usami},\ and\ \citenamefont {Nakamura}}]{Lachance-Quirion2020}%
  \BibitemOpen
  \bibfield  {author} {\bibinfo {author} {\bibfnamefont {D.}~\bibnamefont {Lachance-Quirion}}, \bibinfo {author} {\bibfnamefont {S.~P.}\ \bibnamefont {Wolski}}, \bibinfo {author} {\bibfnamefont {Y.}~\bibnamefont {Tabuchi}}, \bibinfo {author} {\bibfnamefont {S.}~\bibnamefont {Kono}}, \bibinfo {author} {\bibfnamefont {K.}~\bibnamefont {Usami}},\ and\ \bibinfo {author} {\bibfnamefont {Y.}~\bibnamefont {Nakamura}},\ }\bibfield  {title} {\enquote {\bibinfo {title} {Entanglement-based single-shot detection of a single magnon with a superconducting qubit},}\ }\href {https://doi.org/10.1126/science.aaz9236} {\bibfield  {journal} {\bibinfo  {journal} {Science}\ }\textbf {\bibinfo {volume} {367}},\ \bibinfo {pages} {425--428} (\bibinfo {year} {2020})}\BibitemShut {NoStop}%
\bibitem [{\citenamefont {Xu}\ \emph {et~al.}(2023)\citenamefont {Xu}, \citenamefont {Gu}, \citenamefont {Li}, \citenamefont {Weng}, \citenamefont {Wang}, \citenamefont {Li}, \citenamefont {Wang}, \citenamefont {Zhu},\ and\ \citenamefont {You}}]{Xu2023}%
  \BibitemOpen
  \bibfield  {author} {\bibinfo {author} {\bibfnamefont {D.}~\bibnamefont {Xu}}, \bibinfo {author} {\bibfnamefont {X.-K.}\ \bibnamefont {Gu}}, \bibinfo {author} {\bibfnamefont {H.-K.}\ \bibnamefont {Li}}, \bibinfo {author} {\bibfnamefont {Y.-C.}\ \bibnamefont {Weng}}, \bibinfo {author} {\bibfnamefont {Y.-P.}\ \bibnamefont {Wang}}, \bibinfo {author} {\bibfnamefont {J.}~\bibnamefont {Li}}, \bibinfo {author} {\bibfnamefont {H.}~\bibnamefont {Wang}}, \bibinfo {author} {\bibfnamefont {S.-Y.}\ \bibnamefont {Zhu}},\ and\ \bibinfo {author} {\bibfnamefont {J.~Q.}\ \bibnamefont {You}},\ }\bibfield  {title} {\enquote {\bibinfo {title} {Quantum control of a single magnon in a macroscopic spin system},}\ }\href {https://doi.org/10.1103/PhysRevLett.130.193603} {\bibfield  {journal} {\bibinfo  {journal} {Phys. Rev. Lett.}\ }\textbf {\bibinfo {volume} {130}},\ \bibinfo {pages} {193603} (\bibinfo {year} {2023})}\BibitemShut {NoStop}%
\bibitem [{\citenamefont {Baity}\ \emph {et~al.}(2021)\citenamefont {Baity}, \citenamefont {Bozhko}, \citenamefont {Macêdo}, \citenamefont {Smith}, \citenamefont {Holland}, \citenamefont {Danilin}, \citenamefont {Seferai}, \citenamefont {Barbosa}, \citenamefont {Peroor}, \citenamefont {Goldman}, \citenamefont {Nasti}, \citenamefont {Paul}, \citenamefont {Hadfield}, \citenamefont {McVitie},\ and\ \citenamefont {Weides}}]{Baity2021}%
  \BibitemOpen
  \bibfield  {author} {\bibinfo {author} {\bibfnamefont {P.~G.}\ \bibnamefont {Baity}}, \bibinfo {author} {\bibfnamefont {D.~A.}\ \bibnamefont {Bozhko}}, \bibinfo {author} {\bibfnamefont {R.}~\bibnamefont {Macêdo}}, \bibinfo {author} {\bibfnamefont {W.}~\bibnamefont {Smith}}, \bibinfo {author} {\bibfnamefont {R.~C.}\ \bibnamefont {Holland}}, \bibinfo {author} {\bibfnamefont {S.}~\bibnamefont {Danilin}}, \bibinfo {author} {\bibfnamefont {V.}~\bibnamefont {Seferai}}, \bibinfo {author} {\bibfnamefont {J.}~\bibnamefont {Barbosa}}, \bibinfo {author} {\bibfnamefont {R.~R.}\ \bibnamefont {Peroor}}, \bibinfo {author} {\bibfnamefont {S.}~\bibnamefont {Goldman}}, \bibinfo {author} {\bibfnamefont {U.}~\bibnamefont {Nasti}}, \bibinfo {author} {\bibfnamefont {J.}~\bibnamefont {Paul}}, \bibinfo {author} {\bibfnamefont {R.~H.}\ \bibnamefont {Hadfield}}, \bibinfo {author} {\bibfnamefont {S.}~\bibnamefont {McVitie}},\ and\ \bibinfo {author} {\bibfnamefont {M.}~\bibnamefont {Weides}},\ }\bibfield  {title} {\enquote {\bibinfo
  {title} {Strong magnon–photon coupling with chip-integrated {YIG} in the zero-temperature limit},}\ }\href {https://doi.org/10.1063/5.0054837} {\bibfield  {journal} {\bibinfo  {journal} {Appl. Phys. Lett.}\ }\textbf {\bibinfo {volume} {119}},\ \bibinfo {pages} {033502} (\bibinfo {year} {2021})}\BibitemShut {NoStop}%
\bibitem [{\citenamefont {Qin}\ \emph {et~al.}(2018)\citenamefont {Qin}, \citenamefont {H{\"a}m{\"a}l{\"a}inen}, \citenamefont {Arjas}, \citenamefont {Witteveen},\ and\ \citenamefont {van Dijken}}]{Qin2018}%
  \BibitemOpen
  \bibfield  {author} {\bibinfo {author} {\bibfnamefont {H.}~\bibnamefont {Qin}}, \bibinfo {author} {\bibfnamefont {S.~J.}\ \bibnamefont {H{\"a}m{\"a}l{\"a}inen}}, \bibinfo {author} {\bibfnamefont {K.}~\bibnamefont {Arjas}}, \bibinfo {author} {\bibfnamefont {J.}~\bibnamefont {Witteveen}},\ and\ \bibinfo {author} {\bibfnamefont {S.}~\bibnamefont {van Dijken}},\ }\bibfield  {title} {\enquote {\bibinfo {title} {Propagating spin waves in nanometer-thick yttrium iron garnet films: Dependence on wave vector, magnetic field strength, and angle},}\ }\href@noop {} {\bibfield  {journal} {\bibinfo  {journal} {Phys. Rev. B}\ }\textbf {\bibinfo {volume} {98}},\ \bibinfo {pages} {224422} (\bibinfo {year} {2018})}\BibitemShut {NoStop}%
\bibitem [{\citenamefont {Youssef}\ \emph {et~al.}(2025)\citenamefont {Youssef}, \citenamefont {Beaulieu}, \citenamefont {Schlitz}, \citenamefont {Petrosyan}, \citenamefont {Lammel},\ and\ \citenamefont {Legrand}}]{Youssef2025}%
  \BibitemOpen
  \bibfield  {author} {\bibinfo {author} {\bibfnamefont {J.~B.}\ \bibnamefont {Youssef}}, \bibinfo {author} {\bibfnamefont {N.}~\bibnamefont {Beaulieu}}, \bibinfo {author} {\bibfnamefont {R.}~\bibnamefont {Schlitz}}, \bibinfo {author} {\bibfnamefont {D.}~\bibnamefont {Petrosyan}}, \bibinfo {author} {\bibfnamefont {M.}~\bibnamefont {Lammel}},\ and\ \bibinfo {author} {\bibfnamefont {W.}~\bibnamefont {Legrand}},\ }\href {https://arxiv.org/abs/2509.06242} {\enquote {\bibinfo {title} {Low-temperature-compatible iron garnet films grown by liquid phase epitaxy},}\ } (\bibinfo {year} {2025}),\ \Eprint {https://arxiv.org/abs/2509.06242} {arXiv:2509.06242 [cond-mat.mtrl-sci]} \BibitemShut {NoStop}%
\bibitem [{\citenamefont {Will-Cole}\ \emph {et~al.}(2023)\citenamefont {Will-Cole}, \citenamefont {Hart}, \citenamefont {Lauter}, \citenamefont {Grutter}, \citenamefont {Dubs}, \citenamefont {Lindner}, \citenamefont {Reimann}, \citenamefont {Valdez}, \citenamefont {Pearce}, \citenamefont {Monson}, \citenamefont {Cha}, \citenamefont {Heiman},\ and\ \citenamefont {Sun}}]{WillCole2023}%
  \BibitemOpen
  \bibfield  {author} {\bibinfo {author} {\bibfnamefont {A.~R.}\ \bibnamefont {Will-Cole}}, \bibinfo {author} {\bibfnamefont {J.~L.}\ \bibnamefont {Hart}}, \bibinfo {author} {\bibfnamefont {V.}~\bibnamefont {Lauter}}, \bibinfo {author} {\bibfnamefont {A.}~\bibnamefont {Grutter}}, \bibinfo {author} {\bibfnamefont {C.}~\bibnamefont {Dubs}}, \bibinfo {author} {\bibfnamefont {M.}~\bibnamefont {Lindner}}, \bibinfo {author} {\bibfnamefont {T.}~\bibnamefont {Reimann}}, \bibinfo {author} {\bibfnamefont {N.~R.}\ \bibnamefont {Valdez}}, \bibinfo {author} {\bibfnamefont {C.~J.}\ \bibnamefont {Pearce}}, \bibinfo {author} {\bibfnamefont {T.~C.}\ \bibnamefont {Monson}}, \bibinfo {author} {\bibfnamefont {J.~J.}\ \bibnamefont {Cha}}, \bibinfo {author} {\bibfnamefont {D.}~\bibnamefont {Heiman}},\ and\ \bibinfo {author} {\bibfnamefont {N.~X.}\ \bibnamefont {Sun}},\ }\bibfield  {title} {\enquote {\bibinfo {title} {Negligible magnetic losses at low temperatures in liquid phase epitaxy grown {Y}$_{3}${Fe}$_{5}${O}$_{12}$
  films},}\ }\href {https://doi.org/10.1103/PhysRevMaterials.7.054411} {\bibfield  {journal} {\bibinfo  {journal} {Phys. Rev. Mater.}\ }\textbf {\bibinfo {volume} {7}},\ \bibinfo {pages} {054411} (\bibinfo {year} {2023})}\BibitemShut {NoStop}%
\bibitem [{\citenamefont {Haidar}\ \emph {et~al.}(2015)\citenamefont {Haidar}, \citenamefont {Ranjbar}, \citenamefont {Balinsky}, \citenamefont {Dumas}, \citenamefont {Khartsev},\ and\ \citenamefont {Åkerman}}]{Haidar2015}%
  \BibitemOpen
  \bibfield  {author} {\bibinfo {author} {\bibfnamefont {M.}~\bibnamefont {Haidar}}, \bibinfo {author} {\bibfnamefont {M.}~\bibnamefont {Ranjbar}}, \bibinfo {author} {\bibfnamefont {M.}~\bibnamefont {Balinsky}}, \bibinfo {author} {\bibfnamefont {R.~K.}\ \bibnamefont {Dumas}}, \bibinfo {author} {\bibfnamefont {S.}~\bibnamefont {Khartsev}},\ and\ \bibinfo {author} {\bibfnamefont {J.}~\bibnamefont {Åkerman}},\ }\bibfield  {title} {\enquote {\bibinfo {title} {Thickness- and temperature-dependent magnetodynamic properties of yttrium iron garnet thin films},}\ }\href {https://doi.org/10.1063/1.4914363} {\bibfield  {journal} {\bibinfo  {journal} {Journal of Applied Physics}\ }\textbf {\bibinfo {volume} {117}},\ \bibinfo {pages} {17D119} (\bibinfo {year} {2015})}\BibitemShut {NoStop}%
\bibitem [{\citenamefont {Jermain}\ \emph {et~al.}(2017)\citenamefont {Jermain}, \citenamefont {Aradhya}, \citenamefont {Reynolds}, \citenamefont {Buhrman}, \citenamefont {Brangham}, \citenamefont {Page}, \citenamefont {Hammel}, \citenamefont {Yang},\ and\ \citenamefont {Ralph}}]{Jermain2017}%
  \BibitemOpen
  \bibfield  {author} {\bibinfo {author} {\bibfnamefont {C.~L.}\ \bibnamefont {Jermain}}, \bibinfo {author} {\bibfnamefont {S.~V.}\ \bibnamefont {Aradhya}}, \bibinfo {author} {\bibfnamefont {N.~D.}\ \bibnamefont {Reynolds}}, \bibinfo {author} {\bibfnamefont {R.~A.}\ \bibnamefont {Buhrman}}, \bibinfo {author} {\bibfnamefont {J.~T.}\ \bibnamefont {Brangham}}, \bibinfo {author} {\bibfnamefont {M.~R.}\ \bibnamefont {Page}}, \bibinfo {author} {\bibfnamefont {P.~C.}\ \bibnamefont {Hammel}}, \bibinfo {author} {\bibfnamefont {F.~Y.}\ \bibnamefont {Yang}},\ and\ \bibinfo {author} {\bibfnamefont {D.~C.}\ \bibnamefont {Ralph}},\ }\bibfield  {title} {\enquote {\bibinfo {title} {Increased low-temperature damping in yttrium iron garnet thin films},}\ }\href {https://doi.org/10.1103/PhysRevB.95.174411} {\bibfield  {journal} {\bibinfo  {journal} {Phys. Rev. B}\ }\textbf {\bibinfo {volume} {95}},\ \bibinfo {pages} {174411} (\bibinfo {year} {2017})}\BibitemShut {NoStop}%
\bibitem [{\citenamefont {Guo}\ \emph {et~al.}(2022)\citenamefont {Guo}, \citenamefont {McCullian}, \citenamefont {{Chris Hammel}},\ and\ \citenamefont {Yang}}]{Guo2022}%
  \BibitemOpen
  \bibfield  {author} {\bibinfo {author} {\bibfnamefont {S.}~\bibnamefont {Guo}}, \bibinfo {author} {\bibfnamefont {B.}~\bibnamefont {McCullian}}, \bibinfo {author} {\bibfnamefont {P.}~\bibnamefont {{Chris Hammel}}},\ and\ \bibinfo {author} {\bibfnamefont {F.}~\bibnamefont {Yang}},\ }\bibfield  {title} {\enquote {\bibinfo {title} {Low damping at few-k temperatures in {Y}$_3${Fe}$_5${O}$_{12}$ epitaxial films isolated from {Gd}$_3${Ga}$_5${O}$_{12}$ substrate using a diamagnetic {Y}$_3${Sc}$_{2.5}${Al}$_{2.5}${O}$_{12}$ spacer},}\ }\href {https://doi.org/https://doi.org/10.1016/j.jmmm.2022.169795} {\bibfield  {journal} {\bibinfo  {journal} {J. Magn. Magn. Mater.}\ }\textbf {\bibinfo {volume} {562}},\ \bibinfo {pages} {169795} (\bibinfo {year} {2022})}\BibitemShut {NoStop}%
\bibitem [{\citenamefont {Knauer}\ \emph {et~al.}(2023)\citenamefont {Knauer}, \citenamefont {Davídková}, \citenamefont {Schmoll}, \citenamefont {Serha}, \citenamefont {Voronov}, \citenamefont {Wang}, \citenamefont {Verba}, \citenamefont {Dobrovolskiy}, \citenamefont {Lindner}, \citenamefont {Reimann}, \citenamefont {Dubs}, \citenamefont {Urbánek},\ and\ \citenamefont {Chumak}}]{Knauer2023}%
  \BibitemOpen
  \bibfield  {author} {\bibinfo {author} {\bibfnamefont {S.}~\bibnamefont {Knauer}}, \bibinfo {author} {\bibfnamefont {K.}~\bibnamefont {Davídková}}, \bibinfo {author} {\bibfnamefont {D.}~\bibnamefont {Schmoll}}, \bibinfo {author} {\bibfnamefont {R.~O.}\ \bibnamefont {Serha}}, \bibinfo {author} {\bibfnamefont {A.}~\bibnamefont {Voronov}}, \bibinfo {author} {\bibfnamefont {Q.}~\bibnamefont {Wang}}, \bibinfo {author} {\bibfnamefont {R.}~\bibnamefont {Verba}}, \bibinfo {author} {\bibfnamefont {O.~V.}\ \bibnamefont {Dobrovolskiy}}, \bibinfo {author} {\bibfnamefont {M.}~\bibnamefont {Lindner}}, \bibinfo {author} {\bibfnamefont {T.}~\bibnamefont {Reimann}}, \bibinfo {author} {\bibfnamefont {C.}~\bibnamefont {Dubs}}, \bibinfo {author} {\bibfnamefont {M.}~\bibnamefont {Urbánek}},\ and\ \bibinfo {author} {\bibfnamefont {A.~V.}\ \bibnamefont {Chumak}},\ }\bibfield  {title} {\enquote {\bibinfo {title} {Propagating spin-wave spectroscopy in a liquid-phase epitaxial nanometer-thick {YIG} film at millikelvin
  temperatures},}\ }\href@noop {} {\bibfield  {journal} {\bibinfo  {journal} {J. Appl. Phys.}\ }\textbf {\bibinfo {volume} {133}},\ \bibinfo {pages} {143905} (\bibinfo {year} {2023})}\BibitemShut {NoStop}%
\bibitem [{\citenamefont {Schmoll}\ \emph {et~al.}(2025)\citenamefont {Schmoll}, \citenamefont {Voronov}, \citenamefont {Serha}, \citenamefont {Slobodianiuk}, \citenamefont {Levchenko}, \citenamefont {Abert}, \citenamefont {Knauer}, \citenamefont {Suess}, \citenamefont {Verba},\ and\ \citenamefont {Chumak}}]{Schmoll2025}%
  \BibitemOpen
  \bibfield  {author} {\bibinfo {author} {\bibfnamefont {D.}~\bibnamefont {Schmoll}}, \bibinfo {author} {\bibfnamefont {A.~A.}\ \bibnamefont {Voronov}}, \bibinfo {author} {\bibfnamefont {R.~O.}\ \bibnamefont {Serha}}, \bibinfo {author} {\bibfnamefont {D.}~\bibnamefont {Slobodianiuk}}, \bibinfo {author} {\bibfnamefont {K.~O.}\ \bibnamefont {Levchenko}}, \bibinfo {author} {\bibfnamefont {C.}~\bibnamefont {Abert}}, \bibinfo {author} {\bibfnamefont {S.}~\bibnamefont {Knauer}}, \bibinfo {author} {\bibfnamefont {D.}~\bibnamefont {Suess}}, \bibinfo {author} {\bibfnamefont {R.}~\bibnamefont {Verba}},\ and\ \bibinfo {author} {\bibfnamefont {A.~V.}\ \bibnamefont {Chumak}},\ }\bibfield  {title} {\enquote {\bibinfo {title} {Wavenumber-dependent magnetic losses in yttrium iron garnet--gadolinium gallium garnet heterostructures at millikelvin temperatures},}\ }\href {https://doi.org/10.1103/PhysRevB.111.134428} {\bibfield  {journal} {\bibinfo  {journal} {Phys. Rev. B}\ }\textbf {\bibinfo {volume} {111}},\ \bibinfo {pages}
  {134428} (\bibinfo {year} {2025})}\BibitemShut {NoStop}%
\bibitem [{\citenamefont {Serha}\ \emph {et~al.}(2024)\citenamefont {Serha}, \citenamefont {Voronov}, \citenamefont {Schmoll}, \citenamefont {Verba}, \citenamefont {Levchenko}, \citenamefont {Koraltan}, \citenamefont {Davídková}, \citenamefont {Budinská}, \citenamefont {Wang}, \citenamefont {Dobrovolskiy}, \citenamefont {Urbánek}, \citenamefont {Lindner}, \citenamefont {Reimann}, \citenamefont {Dubs}, \citenamefont {Gonzalez-Ballestero}, \citenamefont {Abert}, \citenamefont {Suess}, \citenamefont {Bozhko}, \citenamefont {Knauer},\ and\ \citenamefont {Chumak}}]{Serha2024}%
  \BibitemOpen
  \bibfield  {author} {\bibinfo {author} {\bibfnamefont {R.~O.}\ \bibnamefont {Serha}}, \bibinfo {author} {\bibfnamefont {A.~A.}\ \bibnamefont {Voronov}}, \bibinfo {author} {\bibfnamefont {D.}~\bibnamefont {Schmoll}}, \bibinfo {author} {\bibfnamefont {R.}~\bibnamefont {Verba}}, \bibinfo {author} {\bibfnamefont {K.~O.}\ \bibnamefont {Levchenko}}, \bibinfo {author} {\bibfnamefont {S.}~\bibnamefont {Koraltan}}, \bibinfo {author} {\bibfnamefont {K.}~\bibnamefont {Davídková}}, \bibinfo {author} {\bibfnamefont {B.}~\bibnamefont {Budinská}}, \bibinfo {author} {\bibfnamefont {Q.}~\bibnamefont {Wang}}, \bibinfo {author} {\bibfnamefont {O.~V.}\ \bibnamefont {Dobrovolskiy}}, \bibinfo {author} {\bibfnamefont {M.}~\bibnamefont {Urbánek}}, \bibinfo {author} {\bibfnamefont {M.}~\bibnamefont {Lindner}}, \bibinfo {author} {\bibfnamefont {T.}~\bibnamefont {Reimann}}, \bibinfo {author} {\bibfnamefont {C.}~\bibnamefont {Dubs}}, \bibinfo {author} {\bibfnamefont {C.}~\bibnamefont {Gonzalez-Ballestero}}, \bibinfo {author}
  {\bibfnamefont {C.}~\bibnamefont {Abert}}, \bibinfo {author} {\bibfnamefont {D.}~\bibnamefont {Suess}}, \bibinfo {author} {\bibfnamefont {D.~A.}\ \bibnamefont {Bozhko}}, \bibinfo {author} {\bibfnamefont {S.}~\bibnamefont {Knauer}},\ and\ \bibinfo {author} {\bibfnamefont {A.~V.}\ \bibnamefont {Chumak}},\ }\bibfield  {title} {\enquote {\bibinfo {title} {Magnetic anisotropy and {GGG} substrate stray field in {YIG} films down to millikelvin temperatures},}\ }\href@noop {} {\bibfield  {journal} {\bibinfo  {journal} {npj Spintronics}\ }\textbf {\bibinfo {volume} {2}},\ \bibinfo {pages} {29} (\bibinfo {year} {2024})}\BibitemShut {NoStop}%
\bibitem [{\citenamefont {Serha}\ \emph {et~al.}(2025)\citenamefont {Serha}, \citenamefont {Voronov}, \citenamefont {Schmoll}, \citenamefont {Klingbeil}, \citenamefont {Knauer}, \citenamefont {Koraltan}, \citenamefont {Pribytova}, \citenamefont {Lindner}, \citenamefont {Reimann}, \citenamefont {Dubs}, \citenamefont {Abert}, \citenamefont {Verba}, \citenamefont {Urbánek}, \citenamefont {Suess},\ and\ \citenamefont {Chumak}}]{SERHA2025}%
  \BibitemOpen
  \bibfield  {author} {\bibinfo {author} {\bibfnamefont {R.~O.}\ \bibnamefont {Serha}}, \bibinfo {author} {\bibfnamefont {A.~A.}\ \bibnamefont {Voronov}}, \bibinfo {author} {\bibfnamefont {D.}~\bibnamefont {Schmoll}}, \bibinfo {author} {\bibfnamefont {R.}~\bibnamefont {Klingbeil}}, \bibinfo {author} {\bibfnamefont {S.}~\bibnamefont {Knauer}}, \bibinfo {author} {\bibfnamefont {S.}~\bibnamefont {Koraltan}}, \bibinfo {author} {\bibfnamefont {E.}~\bibnamefont {Pribytova}}, \bibinfo {author} {\bibfnamefont {M.}~\bibnamefont {Lindner}}, \bibinfo {author} {\bibfnamefont {T.}~\bibnamefont {Reimann}}, \bibinfo {author} {\bibfnamefont {C.}~\bibnamefont {Dubs}}, \bibinfo {author} {\bibfnamefont {C.}~\bibnamefont {Abert}}, \bibinfo {author} {\bibfnamefont {R.}~\bibnamefont {Verba}}, \bibinfo {author} {\bibfnamefont {M.}~\bibnamefont {Urbánek}}, \bibinfo {author} {\bibfnamefont {D.}~\bibnamefont {Suess}},\ and\ \bibinfo {author} {\bibfnamefont {A.~V.}\ \bibnamefont {Chumak}},\ }\bibfield  {title} {\enquote {\bibinfo
  {title} {Damping enhancement in {YIG} at millikelvin temperatures due to {GGG} substrate},}\ }\href {https://doi.org/https://doi.org/10.1016/j.mtquan.2025.100025} {\bibfield  {journal} {\bibinfo  {journal} {Materials Today Quantum}\ }\textbf {\bibinfo {volume} {5}},\ \bibinfo {pages} {100025} (\bibinfo {year} {2025})}\BibitemShut {NoStop}%
\bibitem [{\citenamefont {Mitra}\ \emph {et~al.}(2017)\citenamefont {Mitra}, \citenamefont {Cespedes}, \citenamefont {Ramasse}, \citenamefont {Ali}, \citenamefont {Marmion}, \citenamefont {Ward}, \citenamefont {Brydson}, \citenamefont {Kinane}, \citenamefont {Cooper}, \citenamefont {Langridge},\ and\ \citenamefont {Hickey}}]{Mitra2017}%
  \BibitemOpen
  \bibfield  {author} {\bibinfo {author} {\bibfnamefont {A.}~\bibnamefont {Mitra}}, \bibinfo {author} {\bibfnamefont {O.}~\bibnamefont {Cespedes}}, \bibinfo {author} {\bibfnamefont {Q.}~\bibnamefont {Ramasse}}, \bibinfo {author} {\bibfnamefont {M.}~\bibnamefont {Ali}}, \bibinfo {author} {\bibfnamefont {S.}~\bibnamefont {Marmion}}, \bibinfo {author} {\bibfnamefont {M.}~\bibnamefont {Ward}}, \bibinfo {author} {\bibfnamefont {R.~M.~D.}\ \bibnamefont {Brydson}}, \bibinfo {author} {\bibfnamefont {C.~J.}\ \bibnamefont {Kinane}}, \bibinfo {author} {\bibfnamefont {J.~F.~K.}\ \bibnamefont {Cooper}}, \bibinfo {author} {\bibfnamefont {S.}~\bibnamefont {Langridge}},\ and\ \bibinfo {author} {\bibfnamefont {B.~J.}\ \bibnamefont {Hickey}},\ }\bibfield  {title} {\enquote {\bibinfo {title} {Interfacial origin of the magnetisation suppression of thin film yttrium iron garnet},}\ }\href {https://doi.org/https://doi.org/10.1038/s41598-017-10281-6} {\bibfield  {journal} {\bibinfo  {journal} {Sci. Rep.}\ }\textbf {\bibinfo
  {volume} {7}},\ \bibinfo {pages} {11774} (\bibinfo {year} {2017})}\BibitemShut {NoStop}%
\bibitem [{\citenamefont {Kosen}\ \emph {et~al.}(2019)\citenamefont {Kosen}, \citenamefont {van Loo}, \citenamefont {Bozhko}, \citenamefont {Mihalceanu},\ and\ \citenamefont {Karenowska}}]{Kosen2019}%
  \BibitemOpen
  \bibfield  {author} {\bibinfo {author} {\bibfnamefont {S.}~\bibnamefont {Kosen}}, \bibinfo {author} {\bibfnamefont {A.~F.}\ \bibnamefont {van Loo}}, \bibinfo {author} {\bibfnamefont {D.~A.}\ \bibnamefont {Bozhko}}, \bibinfo {author} {\bibfnamefont {L.}~\bibnamefont {Mihalceanu}},\ and\ \bibinfo {author} {\bibfnamefont {A.~D.}\ \bibnamefont {Karenowska}},\ }\bibfield  {title} {\enquote {\bibinfo {title} {Microwave magnon damping in {YIG} films at millikelvin temperatures},}\ }\href {https://doi.org/10.1063/1.5115266} {\bibfield  {journal} {\bibinfo  {journal} {APL Mater.}\ }\textbf {\bibinfo {volume} {7}},\ \bibinfo {pages} {101120} (\bibinfo {year} {2019})}\BibitemShut {NoStop}%
\bibitem [{\citenamefont {Xu}\ \emph {et~al.}(2025)\citenamefont {Xu}, \citenamefont {Horn}, \citenamefont {Jiang}, \citenamefont {Pishehvar}, \citenamefont {Li}, \citenamefont {Rosenmann}, \citenamefont {Han}, \citenamefont {Levy}, \citenamefont {Guha},\ and\ \citenamefont {Zhang}}]{Xu2025}%
  \BibitemOpen
  \bibfield  {author} {\bibinfo {author} {\bibfnamefont {J.}~\bibnamefont {Xu}}, \bibinfo {author} {\bibfnamefont {C.}~\bibnamefont {Horn}}, \bibinfo {author} {\bibfnamefont {Y.}~\bibnamefont {Jiang}}, \bibinfo {author} {\bibfnamefont {A.}~\bibnamefont {Pishehvar}}, \bibinfo {author} {\bibfnamefont {X.}~\bibnamefont {Li}}, \bibinfo {author} {\bibfnamefont {D.}~\bibnamefont {Rosenmann}}, \bibinfo {author} {\bibfnamefont {X.}~\bibnamefont {Han}}, \bibinfo {author} {\bibfnamefont {M.}~\bibnamefont {Levy}}, \bibinfo {author} {\bibfnamefont {S.}~\bibnamefont {Guha}},\ and\ \bibinfo {author} {\bibfnamefont {X.}~\bibnamefont {Zhang}},\ }\bibfield  {title} {\enquote {\bibinfo {title} {Cryogenic hybrid magnonic circuits based on spalled {YIG} thin films},}\ }\href {https://doi.org/10.1063/5.0247663} {\bibfield  {journal} {\bibinfo  {journal} {J. Appl. Phys.}\ }\textbf {\bibinfo {volume} {137}},\ \bibinfo {pages} {023901} (\bibinfo {year} {2025})}\BibitemShut {NoStop}%
\bibitem [{\citenamefont {Legrand}\ \emph {et~al.}(2025)\citenamefont {Legrand}, \citenamefont {Kemna}, \citenamefont {Schären}, \citenamefont {Wang}, \citenamefont {Petrosyan}, \citenamefont {Holder}, \citenamefont {Schlitz}, \citenamefont {Aguirre}, \citenamefont {Lammel},\ and\ \citenamefont {Gambardella}}]{Legrand2025}%
  \BibitemOpen
  \bibfield  {author} {\bibinfo {author} {\bibfnamefont {W.}~\bibnamefont {Legrand}}, \bibinfo {author} {\bibfnamefont {Y.}~\bibnamefont {Kemna}}, \bibinfo {author} {\bibfnamefont {S.}~\bibnamefont {Schären}}, \bibinfo {author} {\bibfnamefont {H.}~\bibnamefont {Wang}}, \bibinfo {author} {\bibfnamefont {D.}~\bibnamefont {Petrosyan}}, \bibinfo {author} {\bibfnamefont {L.}~\bibnamefont {Holder}}, \bibinfo {author} {\bibfnamefont {R.}~\bibnamefont {Schlitz}}, \bibinfo {author} {\bibfnamefont {M.~H.}\ \bibnamefont {Aguirre}}, \bibinfo {author} {\bibfnamefont {M.}~\bibnamefont {Lammel}},\ and\ \bibinfo {author} {\bibfnamefont {P.}~\bibnamefont {Gambardella}},\ }\bibfield  {title} {\enquote {\bibinfo {title} {Lattice-tunable substituted iron garnets for low-temperature magnonics},}\ }\href {https://doi.org/https://doi.org/10.1002/adfm.202503644} {\bibfield  {journal} {\bibinfo  {journal} {Adv. Funct. Mater.}\ }\textbf {\bibinfo {volume} {Early view}},\ \bibinfo {pages} {2503644} (\bibinfo {year}
  {2025})}\BibitemShut {NoStop}%
\bibitem [{\citenamefont {Guo}\ \emph {et~al.}(2023)\citenamefont {Guo}, \citenamefont {Russell}, \citenamefont {Lanier}, \citenamefont {Da}, \citenamefont {Hammel},\ and\ \citenamefont {Yang}}]{Guo2023}%
  \BibitemOpen
  \bibfield  {author} {\bibinfo {author} {\bibfnamefont {S.}~\bibnamefont {Guo}}, \bibinfo {author} {\bibfnamefont {D.}~\bibnamefont {Russell}}, \bibinfo {author} {\bibfnamefont {J.}~\bibnamefont {Lanier}}, \bibinfo {author} {\bibfnamefont {H.}~\bibnamefont {Da}}, \bibinfo {author} {\bibfnamefont {P.~C.}\ \bibnamefont {Hammel}},\ and\ \bibinfo {author} {\bibfnamefont {F.}~\bibnamefont {Yang}},\ }\bibfield  {title} {\enquote {\bibinfo {title} {Strong on-chip microwave photon–magnon coupling using ultralow-damping epitaxial {Y}$_3${Fe}$_5${O}$_{12}$ films at 2 {K}},}\ }\href {https://doi.org/10.1021/acs.nanolett.3c00959} {\bibfield  {journal} {\bibinfo  {journal} {Nano Lett.}\ }\textbf {\bibinfo {volume} {23}},\ \bibinfo {pages} {5055--5060} (\bibinfo {year} {2023})}\BibitemShut {NoStop}%
\bibitem [{\citenamefont {Guguschev}\ \emph {et~al.}(2025)\citenamefont {Guguschev}, \citenamefont {Dubs}, \citenamefont {Blukis}, \citenamefont {Surzhenko}, \citenamefont {Brützam}, \citenamefont {Koc}, \citenamefont {Rhode}, \citenamefont {Berger}, \citenamefont {Richter}, \citenamefont {Berryman}, \citenamefont {Serha},\ and\ \citenamefont {Chumak}}]{Guguschev2025}%
  \BibitemOpen
  \bibfield  {author} {\bibinfo {author} {\bibfnamefont {C.}~\bibnamefont {Guguschev}}, \bibinfo {author} {\bibfnamefont {C.}~\bibnamefont {Dubs}}, \bibinfo {author} {\bibfnamefont {R.}~\bibnamefont {Blukis}}, \bibinfo {author} {\bibfnamefont {O.}~\bibnamefont {Surzhenko}}, \bibinfo {author} {\bibfnamefont {M.}~\bibnamefont {Brützam}}, \bibinfo {author} {\bibfnamefont {R.}~\bibnamefont {Koc}}, \bibinfo {author} {\bibfnamefont {C.}~\bibnamefont {Rhode}}, \bibinfo {author} {\bibfnamefont {K.}~\bibnamefont {Berger}}, \bibinfo {author} {\bibfnamefont {C.}~\bibnamefont {Richter}}, \bibinfo {author} {\bibfnamefont {C.}~\bibnamefont {Berryman}}, \bibinfo {author} {\bibfnamefont {R.~O.}\ \bibnamefont {Serha}},\ and\ \bibinfo {author} {\bibfnamefont {A.~V.}\ \bibnamefont {Chumak}},\ }\href {https://arxiv.org/abs/2508.18101} {\enquote {\bibinfo {title} {Novel diamagnetic garnet-type substrate single crystals for ultralow-damping yttrium iron garnet {Y3Fe5O12} films at cryogenic temperatures},}\ } (\bibinfo {year}
  {2025}),\ \Eprint {https://arxiv.org/abs/2508.18101} {arXiv:2508.18101 [cond-mat.mtrl-sci]} \BibitemShut {NoStop}%
\bibitem [{\citenamefont {Seiden}(1964)}]{Seiden1964}%
  \BibitemOpen
  \bibfield  {author} {\bibinfo {author} {\bibfnamefont {P.~E.}\ \bibnamefont {Seiden}},\ }\bibfield  {title} {\enquote {\bibinfo {title} {Ferrimagnetic resonance relaxation in rare-earth iron garnets},}\ }\href {https://doi.org/10.1103/PhysRev.133.A728} {\bibfield  {journal} {\bibinfo  {journal} {Phys. Rev.}\ }\textbf {\bibinfo {volume} {133}},\ \bibinfo {pages} {A728--A736} (\bibinfo {year} {1964})}\BibitemShut {NoStop}%
\bibitem [{\citenamefont {Das}\ \emph {et~al.}(2023)\citenamefont {Das}, \citenamefont {Mansell}, \citenamefont {Flajšman}, \citenamefont {Yao},\ and\ \citenamefont {van Dijken}}]{Das2023}%
  \BibitemOpen
  \bibfield  {author} {\bibinfo {author} {\bibfnamefont {S.}~\bibnamefont {Das}}, \bibinfo {author} {\bibfnamefont {R.}~\bibnamefont {Mansell}}, \bibinfo {author} {\bibfnamefont {L.}~\bibnamefont {Flajšman}}, \bibinfo {author} {\bibfnamefont {L.}~\bibnamefont {Yao}},\ and\ \bibinfo {author} {\bibfnamefont {S.}~\bibnamefont {van Dijken}},\ }\bibfield  {title} {\enquote {\bibinfo {title} {Perpendicular magnetic anisotropy in {Bi}-substituted yttrium iron garnet films},}\ }\href {https://doi.org/10.1063/5.0184675} {\bibfield  {journal} {\bibinfo  {journal} {J. Appl. Phys.}\ }\textbf {\bibinfo {volume} {134}},\ \bibinfo {pages} {243902} (\bibinfo {year} {2023})}\BibitemShut {NoStop}%
\bibitem [{\citenamefont {Yu}\ \emph {et~al.}(2014)\citenamefont {Yu}, \citenamefont {d'Allivy Kelly}, \citenamefont {Cros}, \citenamefont {Bernard}, \citenamefont {Bortolotti}, \citenamefont {Anane}, \citenamefont {Brandl}, \citenamefont {Huber}, \citenamefont {Stasinopoulos},\ and\ \citenamefont {Grundler}}]{Yu2014}%
  \BibitemOpen
  \bibfield  {author} {\bibinfo {author} {\bibfnamefont {H.}~\bibnamefont {Yu}}, \bibinfo {author} {\bibfnamefont {O.}~\bibnamefont {d'Allivy Kelly}}, \bibinfo {author} {\bibfnamefont {V.}~\bibnamefont {Cros}}, \bibinfo {author} {\bibfnamefont {R.}~\bibnamefont {Bernard}}, \bibinfo {author} {\bibfnamefont {P.}~\bibnamefont {Bortolotti}}, \bibinfo {author} {\bibfnamefont {A.}~\bibnamefont {Anane}}, \bibinfo {author} {\bibfnamefont {F.}~\bibnamefont {Brandl}}, \bibinfo {author} {\bibfnamefont {R.}~\bibnamefont {Huber}}, \bibinfo {author} {\bibfnamefont {I.}~\bibnamefont {Stasinopoulos}},\ and\ \bibinfo {author} {\bibfnamefont {D.}~\bibnamefont {Grundler}},\ }\bibfield  {title} {\enquote {\bibinfo {title} {Magnetic thin-film insulator with ultra-low spin wave damping for coherent nanomagnonics},}\ }\href@noop {} {\bibfield  {journal} {\bibinfo  {journal} {Sci. Rep.}\ }\textbf {\bibinfo {volume} {4}},\ \bibinfo {pages} {6848} (\bibinfo {year} {2014})}\BibitemShut {NoStop}%
\bibitem [{\citenamefont {Schneider}\ \emph {et~al.}(2008)\citenamefont {Schneider}, \citenamefont {Serga}, \citenamefont {Neumann}, \citenamefont {Hillebrands},\ and\ \citenamefont {Kostylev}}]{Schneider2008}%
  \BibitemOpen
  \bibfield  {author} {\bibinfo {author} {\bibfnamefont {T.}~\bibnamefont {Schneider}}, \bibinfo {author} {\bibfnamefont {A.~A.}\ \bibnamefont {Serga}}, \bibinfo {author} {\bibfnamefont {T.}~\bibnamefont {Neumann}}, \bibinfo {author} {\bibfnamefont {B.}~\bibnamefont {Hillebrands}},\ and\ \bibinfo {author} {\bibfnamefont {M.~P.}\ \bibnamefont {Kostylev}},\ }\bibfield  {title} {\enquote {\bibinfo {title} {Phase reciprocity of spin-wave excitation by a microstrip antenna},}\ }\href {https://doi.org/10.1103/PhysRevB.77.214411} {\bibfield  {journal} {\bibinfo  {journal} {Phys. Rev. B}\ }\textbf {\bibinfo {volume} {77}},\ \bibinfo {pages} {214411} (\bibinfo {year} {2008})}\BibitemShut {NoStop}%
\bibitem [{\citenamefont {Sushruth}\ \emph {et~al.}(2020)\citenamefont {Sushruth}, \citenamefont {Grassi}, \citenamefont {Ait-Oukaci}, \citenamefont {Stoeffler}, \citenamefont {Henry}, \citenamefont {Lacour}, \citenamefont {Hehn}, \citenamefont {Bhaskar}, \citenamefont {Bailleul}, \citenamefont {Devolder},\ and\ \citenamefont {Adam}}]{Sushruth2020}%
  \BibitemOpen
  \bibfield  {author} {\bibinfo {author} {\bibfnamefont {M.}~\bibnamefont {Sushruth}}, \bibinfo {author} {\bibfnamefont {M.}~\bibnamefont {Grassi}}, \bibinfo {author} {\bibfnamefont {K.}~\bibnamefont {Ait-Oukaci}}, \bibinfo {author} {\bibfnamefont {D.}~\bibnamefont {Stoeffler}}, \bibinfo {author} {\bibfnamefont {Y.}~\bibnamefont {Henry}}, \bibinfo {author} {\bibfnamefont {D.}~\bibnamefont {Lacour}}, \bibinfo {author} {\bibfnamefont {M.}~\bibnamefont {Hehn}}, \bibinfo {author} {\bibfnamefont {U.}~\bibnamefont {Bhaskar}}, \bibinfo {author} {\bibfnamefont {M.}~\bibnamefont {Bailleul}}, \bibinfo {author} {\bibfnamefont {T.}~\bibnamefont {Devolder}},\ and\ \bibinfo {author} {\bibfnamefont {J.-P.}\ \bibnamefont {Adam}},\ }\bibfield  {title} {\enquote {\bibinfo {title} {Electrical spectroscopy of forward volume spin waves in perpendicularly magnetized materials},}\ }\href {https://doi.org/10.1103/PhysRevResearch.2.043203} {\bibfield  {journal} {\bibinfo  {journal} {Phys. Rev. Res.}\ }\textbf {\bibinfo {volume}
  {2}},\ \bibinfo {pages} {043203} (\bibinfo {year} {2020})}\BibitemShut {NoStop}%
\end{thebibliography}%

\clearpage 

\setcounter{section}{0}
\setcounter{figure}{0}
\setcounter{table}{0}
\setcounter{equation}{0}

\renewcommand{\thesection}{S\arabic{section}}
\renewcommand{\thefigure}{S\arabic{figure}}
\renewcommand{\thetable}{S\arabic{table}}
\renewcommand{\theequation}{S\arabic{equation}}

\begin{center}
\textbf{\large {Spin wave propagation at low temperatures in YIG thin films on YSGG substrates - Supplementary Material}}
\end{center}

\maketitle
\section{Fitting of FMR data}

The FMR measurements were performed using a CryoFMR setup from NanOsc. The system employs lock-in detection, where a small oscillating magnetic field at 490 Hz is applied while sweeping an applied magnetic field at a fixed excitation frequency. The resulting signal corresponds to the derivative of a Lorentzian lineshape, characteristic of the FMR response. The experimental data were fitted using a model consisting of the sum of the derivatives of two Lorentzian functions, incorporating both the real and imaginary components, as described by the following equation:

\begin{equation}
\begin{aligned}
S_{\mathrm{MOD}} &= 
K_{1,1} 
\frac{
\left(\frac{\Delta H_1}{2}\right)\left(H - H_{\mathrm{RES},1}\right)
}{
\left[\left(H - H_{\mathrm{RES},1}\right)^2 + \left(\frac{\Delta H_1}{2}\right)^2\right]^2
}
+ 
K_{2,1} 
\frac{
\left(\frac{\Delta H_1}{2}\right)^2 - \left(H - H_{\mathrm{RES},1}\right)^2
}{
\left[\left(H - H_{\mathrm{RES},1}\right)^2 + \left(\frac{\Delta H_1}{2}\right)^2\right]^2
} \\[6pt]
&\quad +
K_{1,2} 
\frac{
\left(\frac{\Delta H_2}{2}\right)\left(H - H_{\mathrm{RES},2}\right)
}{
\left[\left(H - H_{\mathrm{RES},2}\right)^2 + \left(\frac{\Delta H_2}{2}\right)^2\right]^2
}
+ 
K_{2,2} 
\frac{
\left(\frac{\Delta H_2}{2}\right)^2 - \left(H - H_{\mathrm{RES},2}\right)^2
}{
\left[\left(H - H_{\mathrm{RES},2}\right)^2 + \left(\frac{\Delta H_2}{2}\right)^2\right]^2
}\\
&\quad + H \cdot \text{Slope} + \text{Offset}, 
\end{aligned}
\end{equation}

\noindent where $K_{(1,2),(1,2)}$ are the amplitude coefficients, $H$ is the applied magnetic field, $H_{\mathrm{RES}(1,2)}$ are the resonance fields, and $\Delta H_{(1,2)}$ are the corresponding linewidths. The extracted linewidth is defined as the full width at half maximum (FWHM) of the Lorentzian profile. To identify which of the two fitted peaks exhibited the larger amplitude, the quantities $\sqrt{K_{11}^2 +K_{21}^2}$ and $\sqrt{K_{12}^2+ K_{22}^2}$ were evaluated. The linewidth corresponding to the resonance with the larger amplitude was used for further analysis. Both films show multiple peaks necessitating the double Lorentzian fitting procedure. This is likely due to inhomogeneity across the film arising from the PLD growth.

\clearpage

\section{Extraction of magnetic parameters from FMR data}
FMR measurements were performed as a function of frequency and temperature for both YIG/YSGG and YIG/GGG films. The resonance fields, extracted as described in the previous section, were fitted at each temperature over the frequency range of $2-10$ GHz using the in-plane Kittel formula, which relates the resonance frequency to the applied magnetic field and magnetic material parameters:
\begin{equation}
     f_\mathrm{FMR} = \frac{\gamma\mu_0}{2\pi}\sqrt{(M_\mathrm{eff}+H_\mathrm{k}+H_\mathrm{ext})(H_\mathrm{k}+H_\mathrm{ext})}.
\end{equation}
Here, $ f_\mathrm{FMR}$ is the resonance frequency, $\gamma$ is the gyromagnetic ratio, $M_\mathrm{eff}$ is the effective magnetization defined as $M_\mathrm{eff} = M_\mathrm{s} -H_\mathrm{k}$, $M_\mathrm{s}$ is the saturation magnetization, $H_\mathrm{k}$ is the out-of-plane anisotropy field, and $H_\mathrm{ext}$ is the resonance field. By fitting the experimental FMR data, the temperature dependence of $M_\mathrm{s}$ and $H_\mathrm{k}$ was extracted for both films. The results are presented in Figs.\ S1(a) and S1(b). Across all temperatures, the YIG/GGG film exhibits a higher saturation magnetization than the YIG/YSGG film. At low temperatures, both films develop a small out-of-plane anisotropy, which is likely caused by the mismatch in thermal expansion coefficients between YIG and the respective substrates (GGG or YSGG), resulting in temperature-dependent strain within the films.

\begin{figure*}[htbp]
    \centering
    \includegraphics[width=0.9\linewidth]{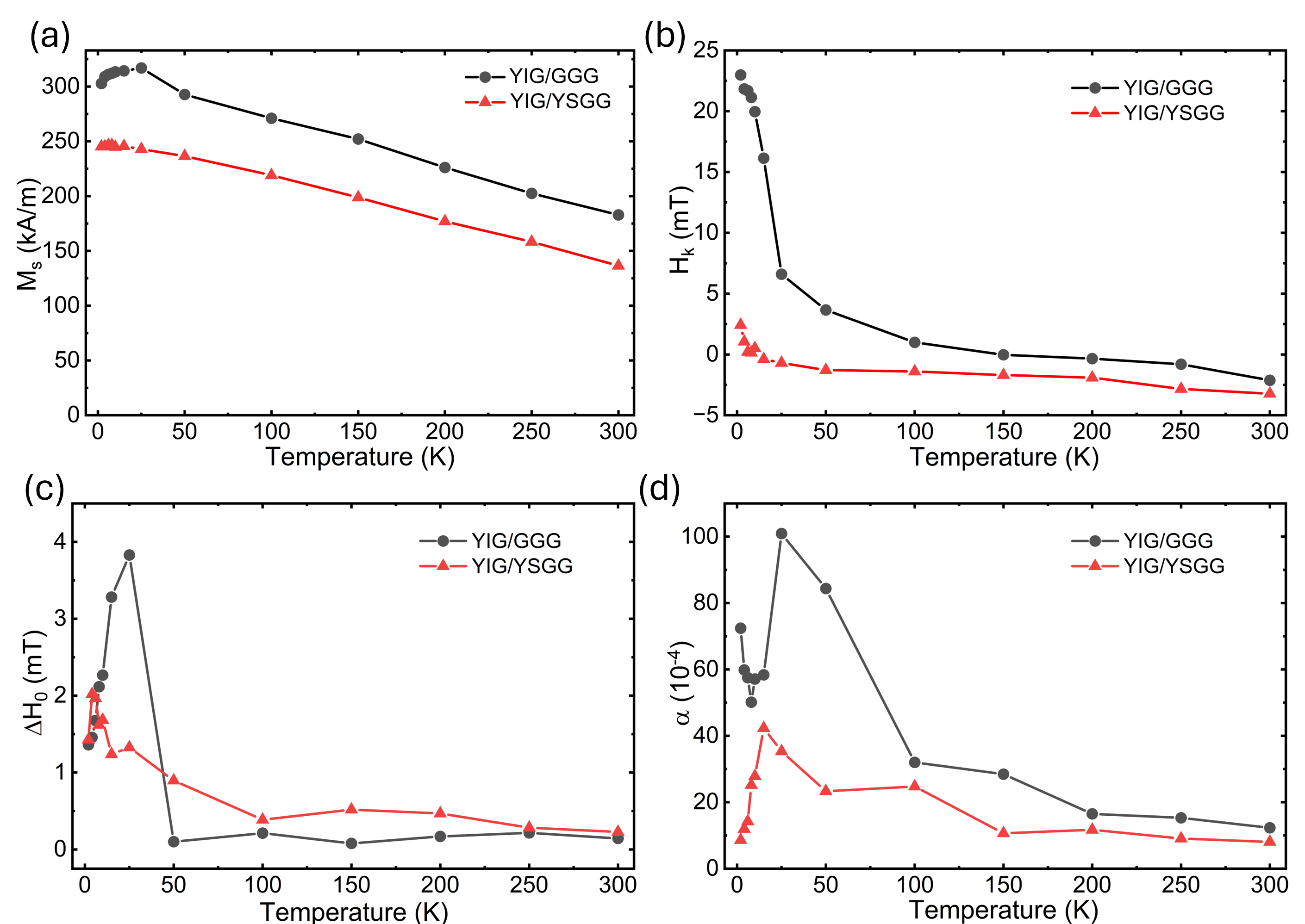}
    \caption{Temperature-dependent parameters extracted from FMR measurements of the YIG/GGG and YIG/YSGG films: (a) saturation magnetization ($M_\mathrm{s}$), (b) out-of-plane magnetic anisotropy field ($H_\mathrm{k}$), (c) inhomogeneous linewidth broadening ($\Delta H_0$), and (d) phenomenological Gilbert damping parameter ($\alpha$).}
    \label{figS1}
\end{figure*}

The inhomogeneous linewidth broadening ($\Delta H_0$) and the phenomenological Gilbert damping parameter ($\alpha$) for both films were determined by fitting the FMR linewidth ($\Delta H$) to the linear relation: 
\begin{equation}
   \Delta H = \Delta H_0 + \frac{4\pi\alpha f_\mathrm{FMR}}{\gamma}.
\end{equation}
The YIG/GGG film shows a pronounced impurity-related peak in $\Delta H_0$ around 25 K, while at higher temperatures and at 2 K its inhomogeneous broadening is comparable to that of the YIG/YSGG film. Both films exhibit similar magnetic damping values at 300 K. However, the YIG/GGG film displays significantly enhanced damping around 25 K. Below this temperature, the damping in the YIG/GGG film decreases to levels similar to those in the YIG/YSGG film, before increasing again at the lowest temperatures measured. In contrast, the YIG/YSGG film exhibits comparable damping values at 2 K and 300 K.

\clearpage

\section{Group velocities}
Group velocities ($\nu_g$) in the YIG/GGG and YIG/YSGG films were extracted from the oscillations in the real part of the S$_{21}$ scattering parameter using the relation: \cite{Yu2014,Qin2018}
\begin{equation}
    \nu_g = \frac{\delta\omega}{\delta k} \approx \frac{2\pi\delta f}{2 \pi / s} = \delta f s,
\end{equation}
where $\delta\omega / \delta k$ defines the group velocity as the rate of change of the angular frequency $\omega$ with respect to the wave vector $k$; $s$ is the antenna separation; and $\delta f$ is the frequency spacing between neighboring peaks in the S$_{21}$ spectra. Since there is some frequency dependence, the first two peaks following the onset of spin-wave propagation in the increasing-frequency direction were used. The extracted group velocities for both films at 2 K and 300 K are shown in Figs.\ S2(a) and S2(b), respectively. 

\begin{figure*}[htbp]
    \centering
    \includegraphics[width=1.0\linewidth]{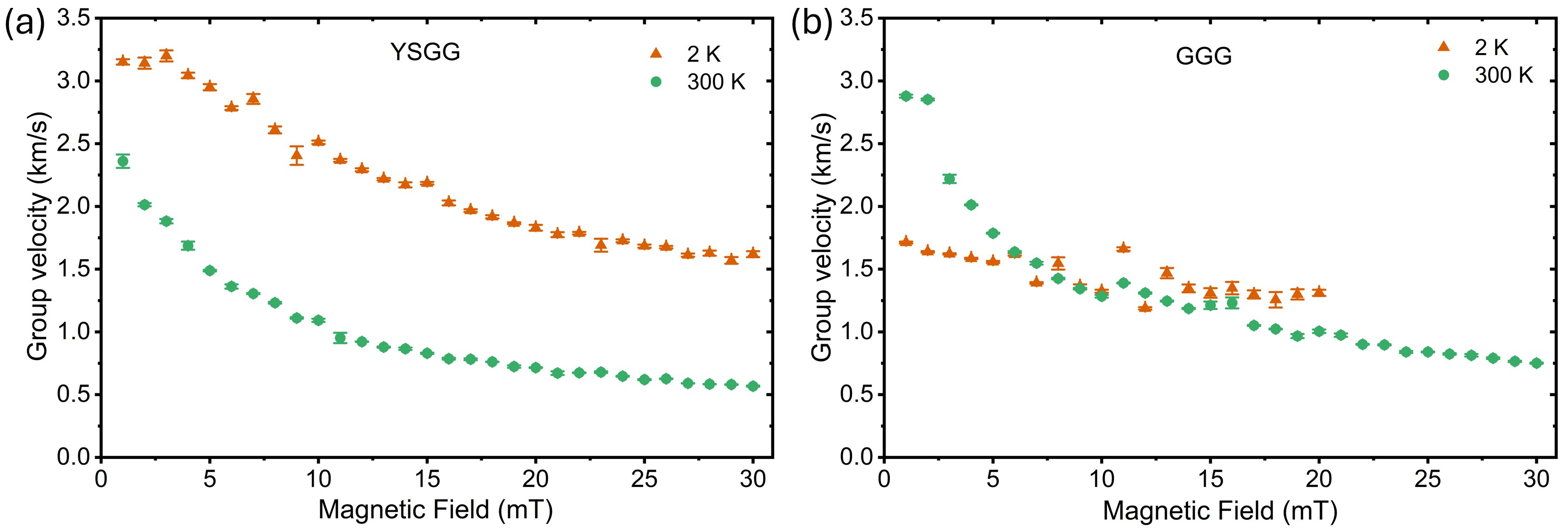}
    \caption{Group velocities as a function of applied magnetic field at 300 K and 2 K for (a) the YIG/YSGG film and (b) the YIG/GGG film. The group velocities were extracted from the data preented in Fig. 3 of the main manuscript.}
    \label{figS3}
\end{figure*}

\clearpage 

\section{Surface roughness}
Figure\ S3 shows atomic force microscopy (AFM) images of the surfaces of both films. The root-mean-square (rms) roughness of the YIG/YSGG film is 0.5 nm, whereas the YIG/GGG exhibits a higher roughness of 1.0 nm. This difference is primarily attributed to the presence of isolated columnar grains in the YIG/GGG film, which protrude above the film surface. 

\begin{figure*}[htbp]
    \centering
    \includegraphics[width=1.0\linewidth]{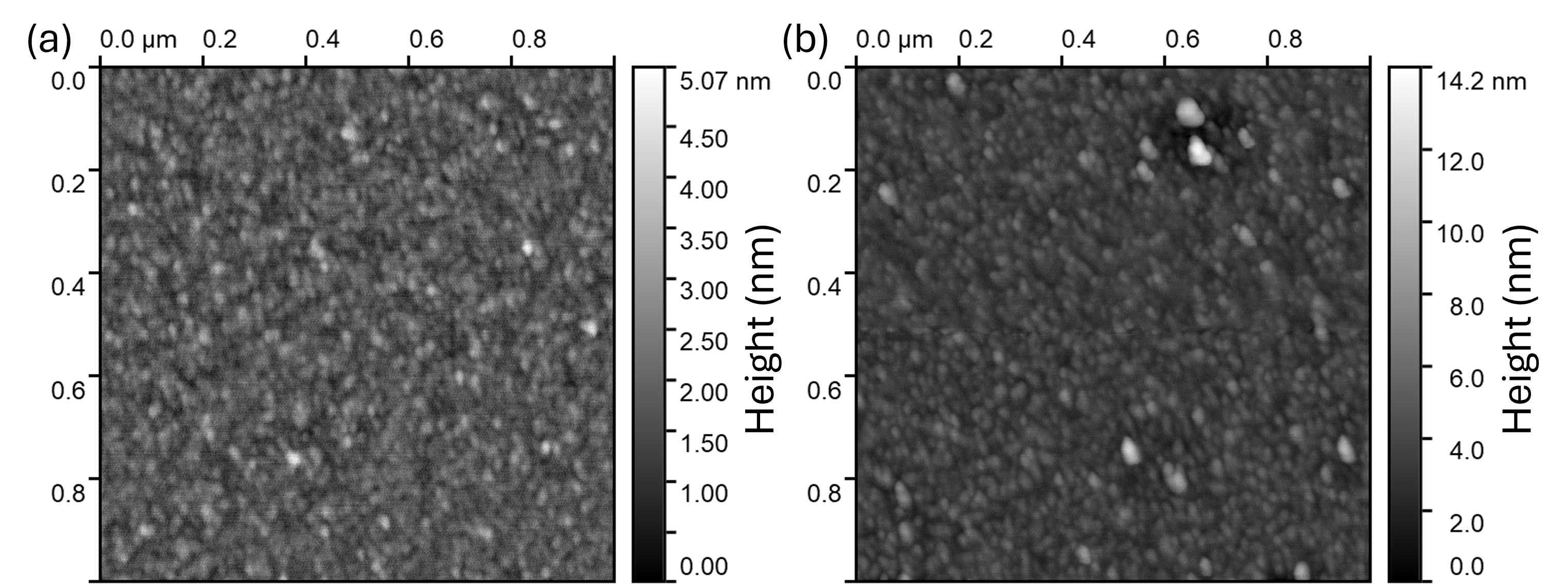}
    \caption{AFM height maps over a 1$\times$1 $\upmu$m$^2$ area for (a) the YIG/YSGG film and (b) the YIG/GGG film.}
    \label{figS4}
\end{figure*}

\clearpage

\end{document}